\documentstyle[preprint,aps]{revtex}

\input{psbox}
\tightenlines
\begin{document}

\draft

%\preprint{Version 2.0}

\title{Numerical signs for a transition in the 2d Random Field Ising
Model at $T=0$}

\author{Carlos Frontera and Eduard Vives}

\address{
Departament d'Estructura i Constituents de la Mat\`{e}ria,
Facultat de F\'{\i}sica, \\Universitat de Barcelona, Diagonal 647,
E-08028 Barcelona, Catalonia, Spain.\\
and \\
Escola Universit\`aria Polit\`ecnica de Matar\'o\\
Av. Puig i Cadafalch 101-111, E-08303 Matar\'o, Catalonia, Spain.}

\date{\today}

\maketitle

\begin{abstract}

Intensive numerical studies of exact ground states of the 2-d ferromagnetic
random field Ising model at $T=0$ with gaussian distribution of fields are
presented. Standard finite size scaling analysis of the data suggests the 
existence of a transition at $\sigma_c =  0.64 \pm 0.08$. Results are 
compared with
existing theories and with the study of metastable avalanches in the same
model.

\end{abstract}


%\narrowtext

The study of systems with quenched disorder has been a challenging problem
since the last 15 years. The interplay between thermal fluctuations and
disorder has a great influence on the existing phase transitions. Many
systems are known to exhibit such phase diagrams highly determined by the
degree of disorder (vacancies, impurities, dislocations, etc.) Among other,
the most typical examples can be found in magnetism, superconductivity,
structural phase transitions, etc. For such systems different models have
been proposed. The Ising model with quenched disorder is one of the simplests
and it has the advantage that the pure model is well known. The disorder can
be of two types: (i) symmetry breaking terms like random-fields or random
magnetic impurities, and (ii) non-symmetry breaking like random-bonds,
vacancies, etc. For all the cases different probability distributions of
disorder have been studied. Here we will focus on the study of the Random
Field Ising Model (RFIM) in two dimensions (2d) with a gaussian distribution
of fields.
Since many years ago there has been a discussion concerning the possibility
that, for such a 2d model with symmetry breaking random fields, it exists or
not order at low temperatures. The initial studies lead to a 
certain controversy: the Imry-Ma \cite{Imry75} argument
suggestes that the lower critical dimension, below which ferromagnetic order
is distroyed, is $d_l \le 2$ with $d=2$ being the limiting case.
Renormalization group expansions\cite{Parisi79} around $d=6$ lead
to the ``dimensional reduction'' argument suggesting that $d_l=3$,
discarding the possibility for ordering in the 2d-RFIM. It has also been 
suggested \cite{Kolomeiskii} that there are new types of order for $d>1$. This
controversy is probably due to the difficulty in balancing the two
ingredients of such models: disorder and thermal fluctuations.

\begin{figure}
\psboxto(0.9\textwidth;0cm){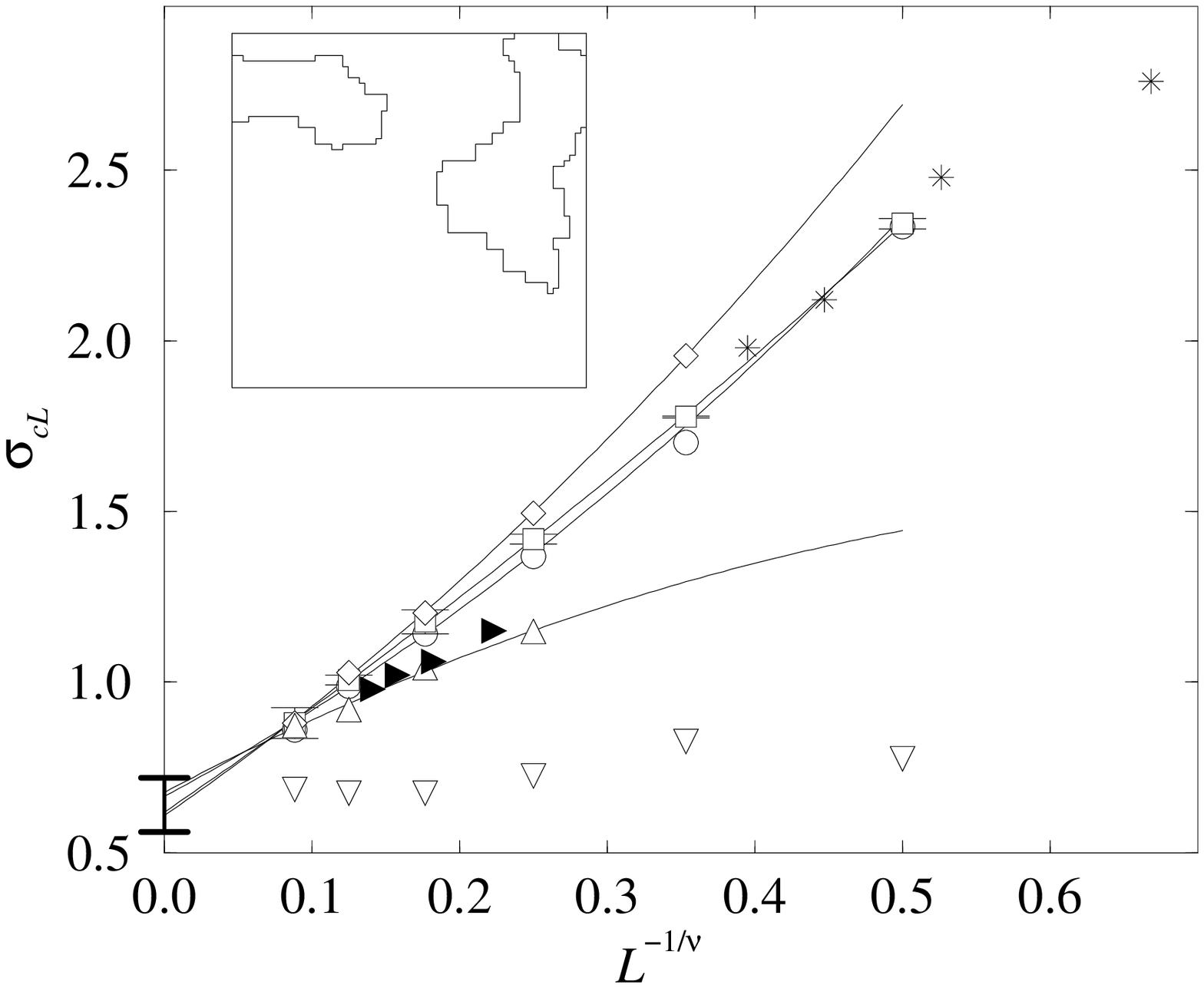}
\caption{$\sigma_{cL}$ versus $L^{-1/\nu}$. The results have been obtained
using MF at zero and higher orders (stars), exact solution
of finite lattices ($\circ$, $\Box$, $\Diamond$, $\bigtriangleup$,
$\bigtriangledown$) and studies of the metastability behaviour (black
triangles) from Ref.[15]. Typical error bars are 
displayed. The inset shows
an example of ground-state of a $L=64$ system with $\sigma=1.0$.
}
\label{FIG1}
\end{figure}

More recently a new approach to disordered systems has been proposed, namely
the study of disordered systems at $T=0$, i.e.~withouth thermal
fluctuations. From a theoretical point of view this simplifies the problem
withouth making it trivial. Moreover, several experimental systems exhibit
phase transitions that can be catalogued into this ``athermal" category: two
examples are ferromagnetism at low temperature under an external magnetic
field \cite{Cote91}, and martensitic transformations\cite{Vives94}. Both
systems present a first-order phase transition that can be crossed by
sweeping a control parameter and are greatly affected by the presence of
quenched disorder. We will concentrate on the study of the 2-d RFIM at $T=0$
for different values of the standard deviation $\sigma$ of the gaussian
distribution of fields. Our goal is to look for signs for the
existence of a phase transition at a certain $\sigma_c$ from a ferromagnetic
ordered state for $\sigma < \sigma_c$ to a disordered state for $\sigma >
\sigma_c$. For the 3d-RFIM at $T=0$, ground state studies
\cite{Ogielski86,Swift97} and  renormalization group arguments
\cite{Newman93} reveal the existence of such phase transition, but to our
knowledge no results for the 2d case have been published. Figure \ref{FIG1}
 summarizes the finite size scaling study
presented in this paper. Data corresponds to estimations of $\sigma_{cL}$
obtained using different methods, as a function of the linear system size
$L^{1/\nu}$ (where $1/\nu=0.5$ is the exponent characterizing the correlation
length divergence). The standard extrapolation to $L \rightarrow \infty$, 
as will be discussed, renders $\sigma_c=0.64 \pm 0.08$ different from zero.

We consider the 2d-RFIM on a $L \times L$ square lattice with periodic
boundary conditions and with the hamiltonian $H = - \sum_{i,j}^{\rm n.n.} S_i
S_j - \sum_{i=1}^{N} S_i h_i$, where $i$ and $j$ are indices sweeping the
full lattice ($i,j=1,\dots,N=L\times L$), the sum refers to
nearest-neighbours (n.n.) pairs, $S_i = \pm 1$ are spin variables and $h_i$
are independent random fields distributed according to the gaussian
probability density with $\langle h \rangle = 0$ and $\langle h^2\rangle =
\sigma^2$. The advantage of using a continuous distribution is that, for
almost any configuration of fields $\lbrace h_i \rbrace$ the ground-sate is
not degenerated. The order parameter is the magnetization of the system
defined as $m= \sum S_i /N$. Because the ground state is unique, thermal
averages are meaningless. Since we are interested in the dependence of the
system properties with the amount of disorder $\sigma$ the only possible
averages with physical meaning are the ensemble averages ($\langle \cdots
\rangle(\sigma)$) performed over different realizations of the random fields
with a certain fixed degree of disorder $\sigma$. Experimentally this has to
be understood as averaging measurements on different samples which have been
prepared with the same amount of disorder.

The zeroth-order mean field (MF) theory was solved 15 years ago
\cite{Aharony78}. A solution with the order parameter
$\langle m\rangle(\sigma) \neq 0$
appears for $\sigma < \sigma_c = \frac{8}{\sqrt{2 \pi}}$, and the phase
transition is continuous. Of course this MF result cannot be
expected to be correct, neither to reflect any dependence on dimensionality.
Moreover, the MF
studies can be extended to higher orders by exactly treating larger and
larger clusters of spins. For thermal phase transitions this is known to
extrapolate to the exact value of the critical temperature. The
first-order approximation is the Bethe approximation which consists in
solving exactly a cluster of a central spin and its four n.n. The method can
be extended to larger clusters. We have found a
continuous phase transition at: $\sigma_c = 2.76, 2.48, 2.12$ and $1.98$ for
clusters of $N=5, 13, 25$ and $41$ spins respectively. These results are
indicated with stars in Fig.~\ref{FIG1} (considering $L=\sqrt{N}$).

A better approach consists in looking for exact ground states by using the
max-flow min-cut theorem \cite{Ogielski86}. We have designed an algorithm
which solves a set of different $\sigma$ values with a minimization time that
grows like $L^4$. We have studied lattices with $L=4$,$8$,$16$,$32$,$64$ and
$128$ and taken averages over $10^5$, $10^4$, $10^4$,$5 \; 10^3$, $10^3$ and
$30$ realizations of random fields respectivelly. The inset in
Fig.~\ref{FIG1} shows a typical example of a ground state for $\sigma=1.0$
and $L=64$. We have focused on the computation of different magnitudes. The
order parameter has been estimated from $\langle |m|\rangle_L(\sigma)$ and
$\sqrt{\langle m^2{\rangle}_L(\sigma)}$. The subscript $L$ indicates that
such quantities will, in general, depend on system size. We have also
measured the susceptibility as $\chi_L(\sigma) = N \left ( \langle
m^2{\rangle}_L - \langle |m|{\rangle}_L^2 \right )$ (which, for large
$\sigma$, tends to $1$ independently of $L$) and the Binder's cumulant
$g_L(\sigma)=1-\langle m^4{\rangle}_L/(3 \langle m^2{\rangle}_L^2)$. Also the
correlation length $\xi_L(\sigma)$ can be computed by fitting an exponential
decay to the spin-spin correlation function.

\begin{figure}
\psboxto(0.9\textwidth;0cm){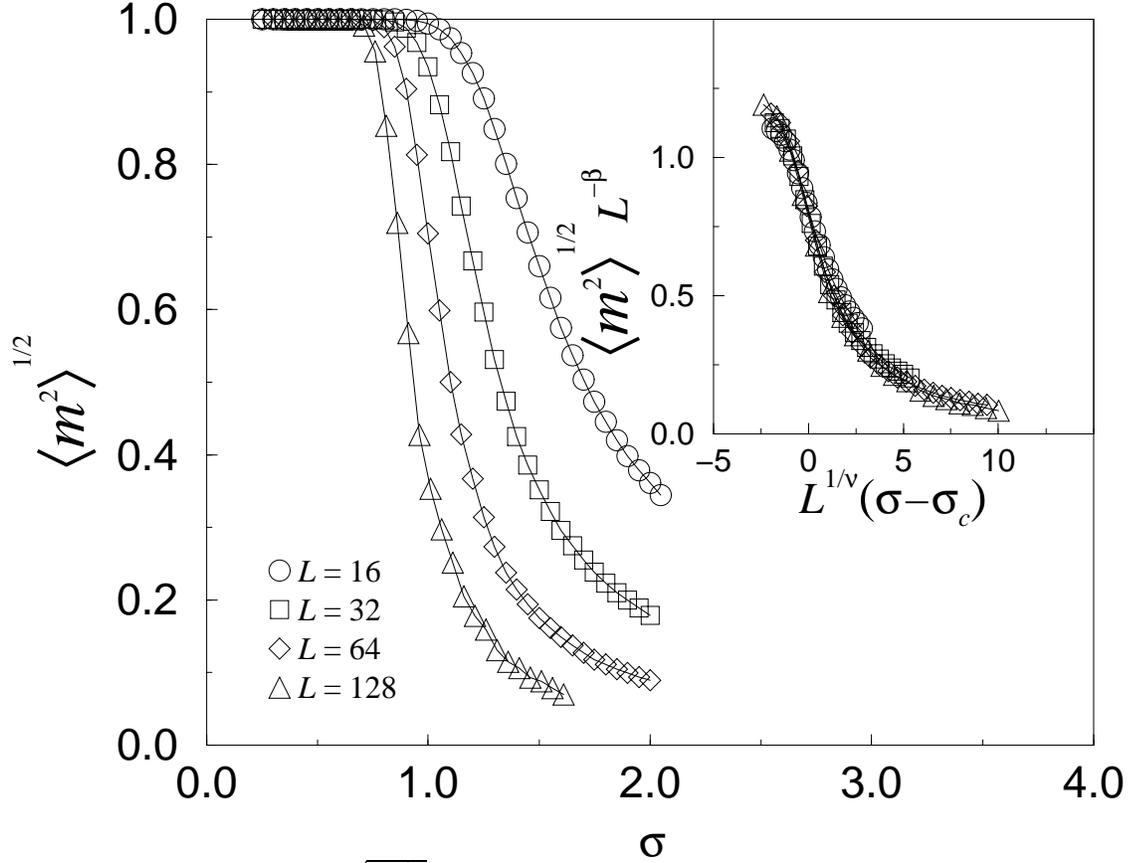}
\caption{Behaviour of $\protect{\sqrt{\langle m^2 \rangle}(\sigma)}$
 for $L=16, 32,
64$ and $128$. The inset shows the same data scaled using $\beta = -0.038$
and $\protect{1/\nu= 0.50}$.}
\label{FIG2}
\end{figure}

\begin{figure}
\psboxto(0.9\textwidth;0cm){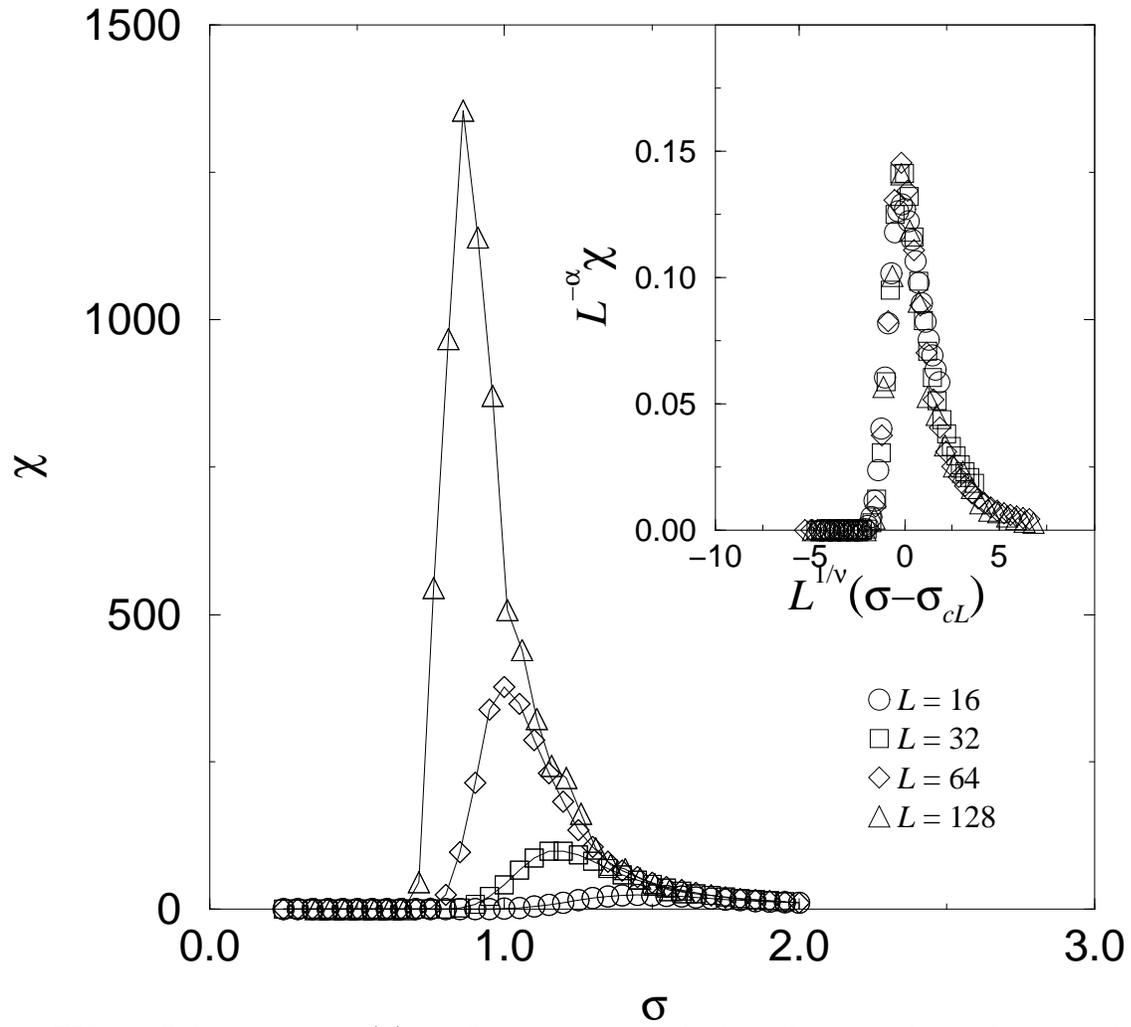}
\caption{Behaviour of $\chi_L(\sigma)$ for $L=16, 32, 64$ and $128$. The
inset shows the same data scaled using $\alpha= 1.89$ and $1/\nu= 0.46$.}
\label{FIG3}
\end{figure}

\begin{figure}
\psboxto(0.9\textwidth;0cm){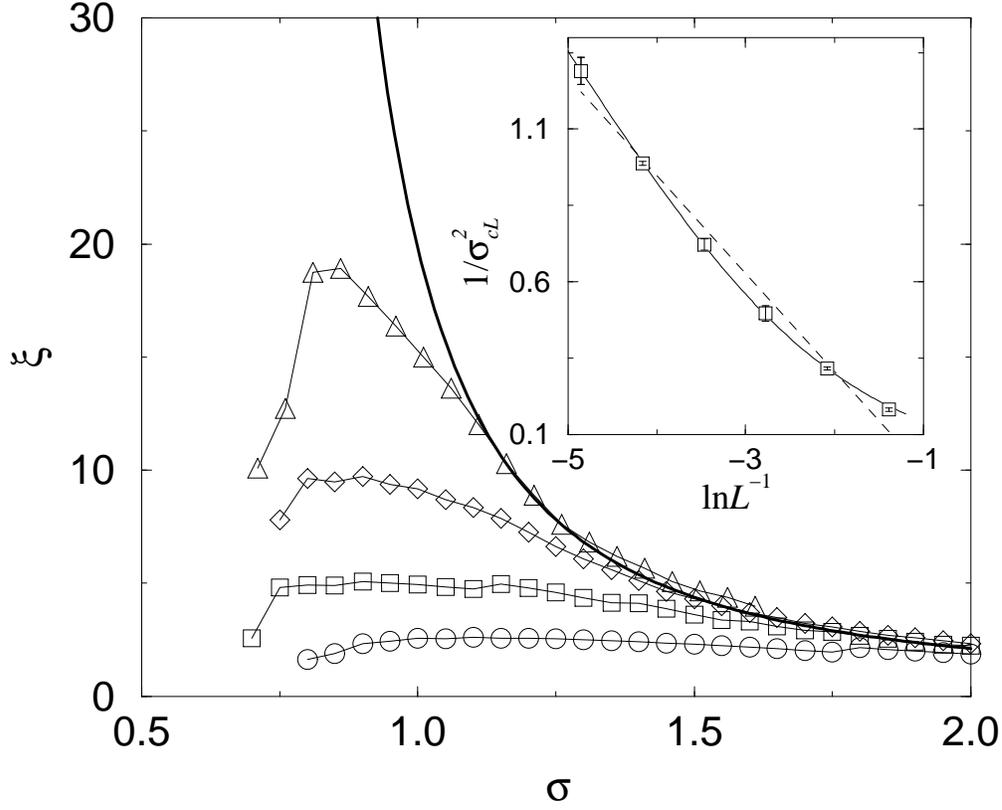}
\caption{Correlation length $\xi_L(\sigma)$ for systems with
$L=16,32,64,128$. The continuous line shows the behaviour of $\xi = A (\sigma
-\sigma_c)^{-\nu} [1 + C (\sigma -\sigma_c)]$ with $\sigma_c= 0.64$ and
$\nu=2$. The inset shows the finite size dependence of $\sigma_{cL}^{-2}$
versus $ln(L^{-1})$. Data is the same as in Fig.~1 ($\Box$). The continuous
line is the standard scaling used in this work and the dashed one is the best
fit of the theory by Binder et al. in Ref.[13].}
\label{FIG4}
\end{figure}

Figure \ref{FIG2} shows the behaviour of $\sqrt{\langle m^2\rangle_L}$ as
a function of $\sigma$ for different system sizes. One can estimate
$\sigma_{cL}$ as the inflection point of a fitted third order polinomium.
Data can be scaled using the standard finite size scaling assumption:
$\sqrt{\langle m^2 \rangle_L} \sim L^\beta \tilde{M} \left [ L^{1/\nu}
(\sigma - \sigma_{cL})\right]$, where $\tilde{M}$ is the corresponding
scaling function. The exponents $\beta$ and $\nu$ can be estimated by fitting
the power-laws $\sqrt{\langle m^2 \rangle}(\sigma=\sigma_{cL})\sim L^\beta$
and $\displaystyle \frac{d\,\sqrt{\langle m^2 \rangle}}{d\,\sigma}
\left ( \sigma = \sigma_{cL} \right ) \sim L^{\beta+1/\nu}$. One gets $\beta=
-0.038\pm 0.009$ and $1/\nu= 0.54 \pm 0.04$. The scaled data is shown in the
inset in Fig.~\ref{FIG2}. A similar scheme can be applied to the study of
$\langle | m |\rangle_L (\sigma)$ rendering $\beta= -0.026\pm 0.017$ and
$1/\nu= 0.54 \pm 0.05 $. Susceptibility $\chi_L(\sigma)$, shown in
Fig.~\ref{FIG3}, exhibits a peak at $\sigma = \sigma_{cL}$ which shifts
and increases when increasing $L$. Data can also be scaled using $\chi_L \sim
L^\alpha \tilde{\chi} \left [ L^{1/\nu} (\sigma - \sigma_{cL})\right]$. Power
law fits to the heigth and curvature of the peak render $\alpha = 1.89 \pm
0.03$ and $1/\nu = 0.46 \pm 0.05$. Scaled data is shown in the inset of
Fig.~\ref{FIG3}. Figure \ref{FIG4} shows the behaviour of $\xi_L(\sigma)$.
The peak gives an independent measure of $\sigma_{cL}$. The continous line is
an estimation of  $\xi(\sigma)$  for $L \rightarrow \infty$ that will be
discused later. Note that the behaviour of the curves is compatible with the
finite size scaling hypothesis, i.e., $\xi_{L}$ follows the behaviour
corresponding to the infinite system up to a certain $\xi_{max} = K L$. (Data
is compatible with $K \simeq 0.08$). A last estimation of $\sigma_{cL}$ can
be obtained from $g_L(\sigma)$. For all the studied sizes $g_L(\sigma)$ takes
a low value for large $\sigma$ and reaches the value $2/3$ at a certain
$\sigma_{cL}$. This estimation is independent of $L$ as suggested in ref.
\cite{Swift97}. We want to note that $\beta \simeq 0$, which means that the
order parameter  increases very fast to $m \simeq 1$ after
the transition. A low-$\sigma$ first-order expansion renders $1-m \sim
10^{-11}$ for $\sigma=0.6$. This fact may also explain why it is very
difficult to measure  $g_L(\sigma)$ with numerical accuracy enough to check
for a crossing point which is the standard procedure to locate the
transition. Note also that $\beta \simeq 0$ and $\alpha \simeq 2$
would suggests that $m$ exhibits a lack of self averaging
\cite{Ferrenberg91}.

The different estimations, as explained in the previous paragraph, of
$\sigma_{cL}$ are plotted in Fig.~\ref{FIG1} in front of $L^{-1/\nu}$. The
open symbols correspond to the estimation from $\langle | m |\rangle_L
(\sigma)$ ($\circ$), $\sqrt{\langle m^2\rangle_L}$ ($\Box$), $\chi_L(\sigma)$
($\Diamond$), $\xi_L(\sigma)$ ($\bigtriangleup$) and $g_L(\sigma)$
($\bigtriangledown$). In order to extrapolate the data to $L \rightarrow
\infty$, we have used the standard expansion for the divergence of $\xi$, up
to second order: $\xi \sim (\sigma-\sigma_{c})^{-\nu}[1+C(\sigma-\sigma_c)]$.
Now, imposing that $\sigma_{cL}$ is determined by the condition $\xi = K L$
one gets: $\sigma_{cL} = \sigma_c + C_1 L^{-1/\nu} + C_2 L^{-2/\nu}$. Such
parabolic fits are also shown in Fig. \ref{FIG1} with continuous lines. The
extrapolated $\sigma_c$ lay all within $\sigma_c = 0.64 \pm 0.08$. To get an
idea of the error margins, we have also fitted the first order expansion
($C_2=0$) leaving $\nu$ free, rendering $\sigma_c=0.65 \pm 0.1$ and $\nu =
1.8 \pm 0.2$; or fixing $\nu=2$ which renders $\sigma_c=0.56 \pm 0.06$.

The existence of this phase transition is in apparent contradiction with
previous results.  It has been proved \cite{Aizenman89} that the RFIM has a
unique Gibbs state in the thermodynamic limit, i.e.  for a given configuration
of the random fields the ground state is unique.  This can be misunderstood
\cite{Nattermann97} as a proof that the ordered phase cannot exist.  When
considering the ensemble of all possible realizations of the random fields
corresponding to a certain value of $\sigma$ it may well be that the
distribution of magnetizations changes from a single peak one (for large
$\sigma$) to a bimodal one for small values of $\sigma$.  Thus the phase
transition we are proposing should be understood as existing in this ensemble
rather than for a single system for which the ground state is unique.It is true
that there is an open question here concerning the size of this ensemble:  in
the thermodynamic limit, is there more than one realization of disorder
compatible with a certain $\sigma$?  We understand that given the discreteness
of the Ising lattice and the continuity of the random fields one can still
consider the existence of such an ensemble.

\begin{figure}
\psboxto(0.9\textwidth;0cm){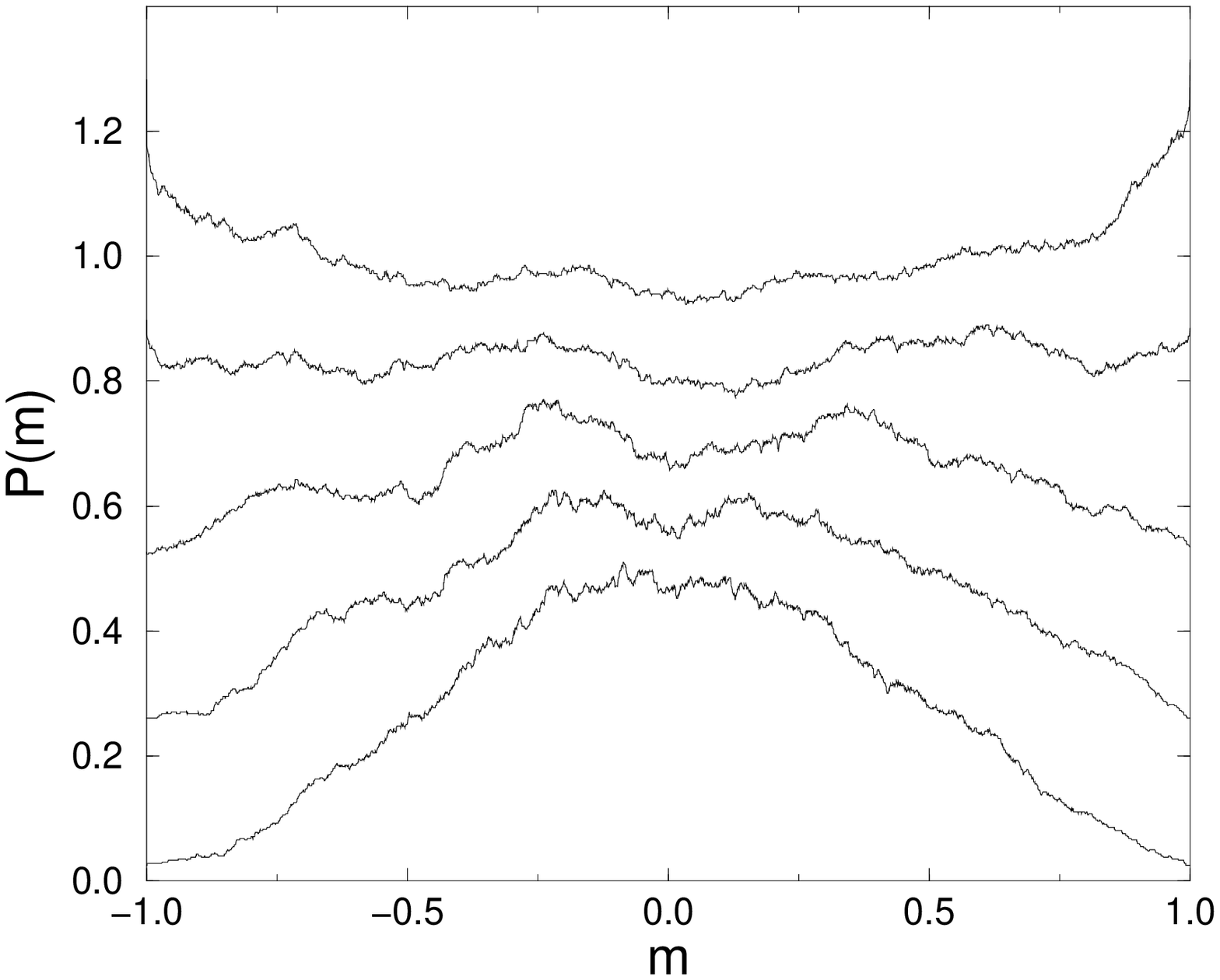}
\caption{Histograms of the magnetization distribution for
a $L=64$ system and for $\sigma= 1.20, 1.15, 1.10, 1.05$ and $1.00$ (from
bottom to top). The distribution evolves from a single peaked one at $\sigma
> \sigma_c$ to a two-peak distribution for $\sigma < \sigma_c$. The curves
are shifted $0.2$ units each to clarify the plot.}
\label{FIG5}
\end{figure}

Another interesting point is the comparison of our
results with studies suggesting an exponential divergence of the correlation
length $\xi \sim \exp\{-B/(\sigma-\sigma_c)^\tau\}$. On the basis of the 
study of the interfaces separating regions with $m>0$ and $m<0$ 
Binder\cite{Binder83} derived a theory with $\tau=2$ and $\sigma_c=0$. We have
tested the validity of this theory by studying the corresponding finite size
scaling hypothesis \cite{Binder83,Bray85}: $1/\sigma_{cL}^2 \propto ln 
(L^{-1})$. The
inset in Fig. \ref{FIG4} shows the comparison of this behaviour with the
standard one we propose in this work on top of the data corresponding to the
estimations of $\sigma_{cL}$ from the inflection point in $\sqrt{\langle
m^2\rangle_L}$ (both fits have two free parameters). The standard theory 
works better. We have also tested
that, assuming the standard hypothesis ($\xi_L \sim \left ( \left |
\frac{\sigma - \sigma_{cL}}{\sigma_{cL}} \right | L^{1/\nu } +
A \right )^{-\nu}$), the finite size scaling of $\xi_L$ is better than using
Binder's hypothesis. Moreover, Binder's theory proposes that for large enough
systems the configurations with total $m \simeq 0$ will
be more and more frequent. We have not observed the existence of many of such
``slab" configurations but found ground-states with closed domains like that
in the inset of Fig.~\ref{FIG1}. Figure \ref{FIG5} shows the probability
$P(m)$ obtained from the computations of a very large number of exact
ground-states for a system with $L=64$. Clearly, for $\sigma < \sigma_c$, the
configurations with $m \simeq 0$ have much less probability than the
configurations with $m \simeq 1$. The reason for the failure of Binder's theory
could be that, in order to perform the thermodynamic limit he uses a very 
anisotropic system with open boundary conditions.

Our data can be compared with the studies of the evolution of the
RFIM at $T=0$, obbeying a local relaxation dynamics. It has been found that
when sweeping the external field the system evolves by avalanches between
metastable states. At a certain degree of disorder $\sigma_c$ the
distribution of avalanches becomes critical. In Fig.~\ref{FIG1} we show the
values of the $\sigma_{cL}$ corresponding to the 2d case from Ref.
\cite{Vives95}. The behaviour is very similar to the equilibirum data.
Different extrapolations to $L \rightarrow \infty$ have been reported
($\sigma_c= 0.75 \pm 0.03$ \cite{Vives95}, $\sigma_c=0.54 \pm 0.04$
\cite{Sethna97}) but all are close to the equilibrium one. Concerning the
exponents, for the metastable studies the exponent $\beta$ has also been
found to be very small, while previous reported values for $\nu$ are $1.6 \pm
0.1$ \cite{Vives95} and $5.3 \pm 1.4$ \cite{Sethna97}. Therefore we suggest
that the metastability phenomena found in the out-of-equilibrium studies
might be associated with a real underlying equilibrium phase transition at
$\sigma < \sigma_c$ for zero external field. It should be mentioned that in
the context of these out of equilibrium phase transitions Sethna and 
collaborators have
\cite{Sethna97} proposed a theory with exponential divergence of $\xi$ with
$\tau=1$ and $\sigma_c=0.42 \pm 0.04$. Our data is not consistent with 
such theory. 
If we perform a fit in the evolution of $\sigma_{cL}(L)$ leaving $\sigma_c$ and
$\tau$ free we get $\tau=0.6$ and $\sigma_c=0.25$. We can obtain still a good 
fit (and good scalings) by taking $\tau=1$ and $\sigma_c=0$, although we cannot
provide any physical explanation for such a behaviour. 
 We finally want to point out
that the phase transition we have found at $T=0$ may also be related to the
change in the type of growth found at $\sigma=0.33$ in the studies of the
depinning transition in the same model\cite{Ji91}.

In conclusion, we have presented a finite size scaling analysis of numerical
data for systems up to $L=128$, that suggests that the RFIM with a gaussian 
distribution of fields, at
$T=0$ exhibits, a phase transition at $\sigma_c=0.64 \pm 0.08$. The ensemble
average of the magnetization changes from $\langle m \rangle = 0$ for $\sigma
> \sigma_c$ to a state with $\langle m \rangle \neq 0$ for $\sigma <
\sigma_c$. The transition is characterized by the exponents $1/ \nu=0.5\pm
0.05$, $\beta= -0.03 \pm 0.02$ and $\alpha=1.89 \pm 0.03$. The possibility of a
exponential divergence $\xi\sim \exp(B/\sigma)$ cannot be excluded.

We acknowledge finantial support from CICyT project number (mat95-0504), CRAY
and FCR. C.F. also acknowledges the Comissionat per a Universitats i Recerca
(Generalitat de Catalunya) for finantial support.

\newpage

\begin{references}
\bibitem{Binder83} K.Binder, Z.Phys. {\bf B 50}, 343 (1983).
He studies a RFIM with the $\pm h$ distribution but we
think that his arguments can also be applied to the present gaussian
distribution.
\end{references}
\end{document}